%% file: main.tex
\documentclass[manuscript,screen,authorversion,nonacm]{acmart} 

\AtBeginDocument{%
  \providecommand\BibTeX{{%
    \normalfont B\kern-0.5em{\scshape i\kern-0.25em b}\kern-0.8em\TeX}}}

\setcopyright{acmlicensed}
\copyrightyear{2024}
\acmYear{2024}

\usepackage[utf8]{inputenc}
\usepackage[binary-units]{siunitx}
\usepackage{balance}
\usepackage{nicefrac} 
\usepackage{acronym}
\usepackage{xspace}

\usepackage{mathtools}
\DeclarePairedDelimiter\bra{\langle}{\rvert}
\DeclarePairedDelimiter\ket{\lvert}{\rangle}
\DeclarePairedDelimiterX\braket[2]{\langle}{\rangle}{#1\,\delimsize\vert\,\mathopen{}#2}

\usepackage{caption}
\usepackage{subcaption}
\usepackage{multirow}
\usepackage{booktabs}

\usepackage{stmaryrd} 

\newcommand{\fakeparagraph}[1]{\vspace{.5mm}\textbf{#1.}}

\begin{document}

\title{Variational Quantum Algorithms for Combinatorial Optimization}

\author{Daniel F. Perez-Ramirez}
\email{daniel.perez@ri.se -- dfpr@kth.se}
\orcid{0000-0002-1322-4367}
\affiliation{%
  \institution{RISE Computer Science}
  \country{Sweden}
}
\affiliation{%
  \institution{KTH Royal Institute of Technology}
  \city{Kista}
  \state{Stockholm}
  \country{Sweden}
}

\renewcommand{\shortauthors}{Perez-Ramirez}

\begin{abstract}
    The promise of quantum computing to address complex problems requiring high computational resources has long been hindered by the intrinsic and demanding requirements of quantum hardware development.
    Nonetheless, the current state of quantum computing, denominated Noisy Intermediate-Scale Quantum (NISQ) era, has introduced algorithms and methods that are able to harness the computational power of current quantum computers with advantages over classical computers (referred to as \emph{quantum advantage}). Achieving quantum advantage is of particular relevance for the combinatorial optimization domain, since it often implies solving an NP-Hard optimization problem. Moreover, combinatorial problems are highly relevant for practical application areas, such as operations research, or resource allocation problems.
    Among quantum computing methods, Variational Quantum Algorithms (VQA) have emerged as one of the strongest candidates towards reaching practical applicability of NISQ systems.
    This paper explores the current state and recent developments of VQAs, emphasizing their applicability to combinatorial optimization. We identify the Quantum Approximate Optimization Algorithm (QAOA) as the leading candidate for these problems. Furthermore, we implement QAOA circuits with varying depths to solve the MaxCut problem on Erdős–Rényi graphs with 10 and 20 nodes, demonstrating the potential and challenges of using VQAs in practical optimization tasks. We release our code, dataset and optimized circuit parameters under \url{https://github.com/DanielFPerez/VQA-for-MaxCut}.
\end{abstract}




\maketitle

\section{Introduction}

\acp{COP} are a class of mathematical problems where the objective is to find the best solution from a finite set of possible solutions~\cite{korte2011combinatorial, papadimitriou2013combinatorial}. These problems are characterized by an objective function that needs to be optimized (either minimized or maximized) within a discrete, but often large, configuration space. \acp{COP} are highly relevant across various fields, including logistics, scheduling, operations research, and particularly resource allocation in networking~\cite{Bengio2021mlforcop, vesselinova2020learning}. The importance of COPs in resource allocation stems from the need to optimize the use of limited resources to maximize efficiency and performance. In particular, many resource allocation problems can be traced back to traditional \acp{COP}. Many of these problems are NP-hard~\cite{korte2011combinatorial}, meaning that the time required to find an exact solution increases exponentially with the size of the input. This intrinsic complexity makes them challenging to solve using classical computational methods. Although data-driven machine learning methods have proven to reduce the optimality gap over traditional heuristics algorithms~\cite{Bengio2021mlforcop, vesselinova2020learning}, the complex and challenging nature of \acp{COP} position them as a highly relevant research area.

Quantum computing has emerged as a promising frontier for addressing these complex problems. 
Quantum computing offers the potential of revolutionizing the way we approach these complex problems through quantum advantage and quantum supremacy~\cite{arute2019quantum, zhong2020quantum}. Quantum advantage refers to the scenario where a quantum computer can solve a problem faster than the best-known classical algorithms, while quantum supremacy is when a quantum computer can solve a problem that is practically unsolvable by classical computers~\cite{cerezo2021variational}. However, current quantum hardware is limited, falling into the category of Noisy Intermediate-Scale Quantum (NISQ) era devices~\cite{Preskill2018quantumcomputingin}. These devices, characterized by their susceptibility to errors and limited qubit counts, are not yet fully capable of achieving the long-term goal of fault-tolerant quantum computation. Nonetheless, \ac{NISQ} devices offer a platform for developing and testing quantum algorithms proven to achieve quantum advantages~\cite{cerezo2021variational}.

\acp{VQA} have emerged as the most promising candidates for leveraging NISQ devices to tackle complex problems~\cite{cerezo2021variational}, including combinatorial optimization. \acp{VQA} operate by using parameterized quantum circuits, whose parameters are optimized using classical algorithms to minimize a cost function. This hybrid approach allows for the combination of quantum computational power with classical optimization techniques. Despite their potential, \ac{VQA} face significant challenges such as extensive hyper-parameter tuning and the phenomenon of \ac{BP}, where gradients become exceedingly small, making training difficult~\cite{mcclean2018barren}. The \ac{QAOA} is a prominent example of a \ac{VQA} designed specifically for \acp{COP}~\cite{farhi2014quantum}. \ac{QAOA} alternates between applying problem-specific and mixing unitary operations to a quantum state, iteratively improving the solution.

In this paper, we explore the application of \ac{VQA} to combinatorial optimization problems, with a focus on \ac{QAOA}. We provide a description of \acp{VQA}, discuss the operational principles of \ac{QAOA}, and implement \ac{QAOA} for solving the MaxCut problem -- a classic NP-hard problem in graph theory. By investigating the implementation and performance of \ac{QAOA} on the MaxCut problem, we aim to contribute to the understanding and development of practical quantum algorithms for combinatorial optimization.

\section{Variational Quantum Algorithms}

\begin{figure}
    \centering
    \includegraphics[width=0.8\textwidth]{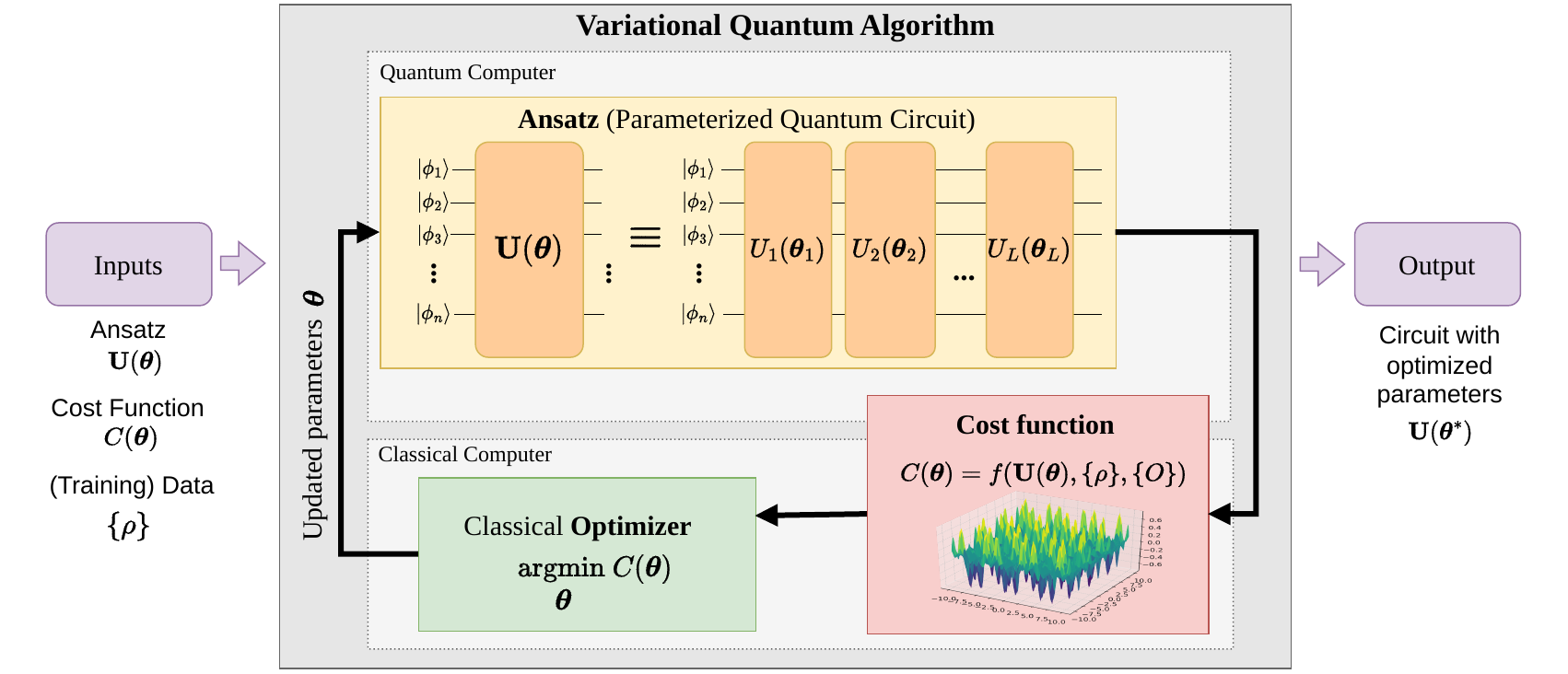}
    \caption{
    \textbf{Components of a Variational Quantum Algorithm}. 
    }
    \label{fig:vqa-overview}
\end{figure}

This section delves into the components, applications, and challenges associated with \ac{VQA}, providing a comprehensive understanding of their operational framework and potential.

\acp{VQA} are a class of quantum algorithms that leverage both quantum and classical computational resources to solve optimization problems. These algorithms iteratively optimize a parameterized quantum circuit using a classical optimizer to minimize a cost function that encodes the problem of interest. \acp{VQA} are particularly suited for near-term quantum devices, as they can mitigate noise and other hardware limitations through their hybrid approach.

\subsection{Components of a VQA}
A \ac{VQA} consists mainly of four different components as observed in Fig.~\ref{fig:vqa-overview}: a cost function, a parameterized quantum circuit, a classical optimizer and possible a set of training data.

The \textbf{\textit{cost function}} $C(\boldsymbol{\theta})$ represent the hyper-surface to minimize for finding the solution to the problem at hand. In general, it is a function that depends on a quantum circuit $U$, a set of input training data $\rho$ with observables $O$: $C(\boldsymbol{\theta})=f(U(\boldsymbol{\theta}), \{\rho\}, \{O\})$. It is obtained by performing measurements on a \ac{QC}, and is akin a loss function in the traditional \ac{ML} setting (trainable and measurable function).    

The parameterized quantum circuit, also called an \textbf{\textit{Ansatz}} $U(\boldsymbol{\theta})$, consists of a series of unitary transformations with learnable parameters $\boldsymbol{\theta}$ that receive an input quantum state and produce as output a bitstring after a measurement operation to the quantum circuit. 
However, current \ac{NISQ} hardware is prone to errors, has short decoherence time\footnote{The time a qubit keeps its state without being distorted from environment noise.} and limited qubits, which ultimately limits the depth of the Ansatz.
There are several standard Ansätze used in practice, both problem specific and problem agnostic, akin to neural network model architectures. 
Among those, the quantum alternating operator Ansatz used in \ac{QAOA} is of specific relevance for \ac{COP}. This Ansatz takes the input quantum state $\ket{\phi}$ and transforms it through $p$ iterations of alternating transformations of a problem unitary and a mixer unitary. 
Other interesting approaches for Ansätze include optimizing not only the parameters of the Ansatz, but its structure (the operators that form it)~\cite{grimsley2019adaptive}, and hybrid approaches that offload some of the quantum operations to classical devices~\cite{bharti2021quantassistsim}. 

The \textbf{\textit{optimizer}} is responsible of training the parameters $\boldsymbol{\theta}$ while minimizing the cost function. For such purposes, the gradient of cost function w.r.t. the parameters can be evaluated through the parameter-shift rule~\cite{schuld2019parametershift} and the chain rule. The parameter-shift rule is a method that resembles finite differences and calculates the gradient by shifting a given parameter by a fixed precision term~\cite{schuld2019parametershift,cerezo2021variational}. 
Similar to deep learning, cost functions are expected to have many local minima. Given that we must resort to statistical estimates of the gradients, traditional gradient descent optimizers from the classical optimization and \ac{ML} domain are deployed for parameter update. However, in contrast to \ac{ML} applications where one input is processed once by epoch by the \ac{ML} model, a \ac{VQA} model (or any \ac{QC} for that regard) requires multiple measurements of the same input. Hence, other optimizers are specifically designed to reduce the number of quantum measurements. 
Further methods explore the concept of information geometry through metric tensors, through meta-learning by training a \ac{ML} to do the parameter updates~\cite{wilson2021optimizing}, or through gradient-free methods that lower the computational demands by using only approximations of the gradients~\cite{spall1992multivariate}.   

\subsection{Application of VQAs}

A key highlight of \acp{VQA} is their task-oriented programming nature -- similar to deep \acp{NN}, \acp{VQA} can be deployed for a wide variety of tasks. 

The most evident application of \ac{VQA} relates to its connection with physical quantum systems: finding energy levels (ground or exciting states) of atoms or many-body problems~\cite{abrams1999quant4chem, peruzzo2014variational}. This is done, in general, by computing expectations of the problem Hamiltonian, which describes the total energy of the system. Beyond the static case, \ac{VQA} can also be used to estimate the evolution of a dynamic quantum system~\cite{li2017quant4dynamsys}.

Perhaps the most known use-case for deploying \ac{QC} to achieve exponential gains and outperform classical computers (i.e., to achieve \textit{quantum supremacy}) is that of mathematical applications~\cite{shor1994shor, harrow2009hhl}. While there are several quantum algorithms already proposed for the fault-tolerant quantum era\footnote{\acp{QC} that can continue functioning correctly even in the presence of errors and faults.}, \acp{VQA} present an alternative to achieve heuristical scalings of such algorithms while still operating under the constraints of \ac{NISQ}. Applications include solving (non)linear systems of equations and factorization~\cite{shor1994shor, harrow2009hhl}.

As aforementioned, \acp{VQA} have also been applied for \ac{COP}. \ac{QAOA} is the most known approach for this purpose, and it has found application for solving both constraint satisfaction problems and MaxCut problems~\cite{farhi2014quantum, zhu2022adaptive, khairy2020learning}. 
Finally, \acp{VQA} have also been applied for \ac{QML} to learn patterns in quantum data~\cite{schuld2019quantum, cerezo2021variational}. Examples can be found through classifiers by exploiting the Hilbert space produced by the quantum state as a feature space, quantum Autoencoders for compressing quantum data,  and generative models~\cite{cerezo2021variational}.  

\subsection{Challenges \& Outlook of VQAs}
\ac{VQA} face several challenges in terms of their trainability, accuracy and efficacy. 
On their \textit{\textbf{trainability}}, perhaps the most known and concerning limitation is the existence of so-called \ac{BP}: a phenomenon in \ac{VQA} systems where the magnitude of the partial derivatives of the cost function (required for computing the gradient and subsequent parameter update) exponentially vanish with increasing quantum circuit depth~\cite{, arrasmith2021effect, mcclean2018barren}. 
This implies that \ac{VQA} need an exponentially increasing precision to counter the effect of \ac{BP}~\cite{cerezo2021variational}. 
Even for shallow circuits, the existence of \ac{BP} is proven to be present and is cost function-dependent. However, the authors describe some approaches recently develop to avoid \ac{BP}: the choice of the Ansatz and the strategy to initialize the Ansatz parameters (akin to \ac{NN} weight initialization)~\cite{grant2019initialization}. The effect and influence of initialization for the occurrence of \ac{BP} was first identified by \citet{zhou2020paraminitforQAOA} for the case of \ac{QAOA}.
    
The \textit{\textbf{efficiency}} issue in \ac{VQA} relates to calculating expectation values in the circuits (used for computing the cost functions), i.e., the number of measurements required at the circuit's output. This is also referred to as \textit{measurement frugality}. Approaches to mitigate this issue are mainly centered towards making measurements that are projected onto the eigenbasis of the operator (circuit unitary). Due to the high computational cost and possible intractability of finding such eigenbasis, a common approach is to decompose the operator into a sum of simpler Pauli operators~\cite{cerezo2021variational, khairy2020learning}. 

Finally, the issue of \textbf{\textit{accuracy}} relates to handling and accounting for hardware noise. One possible alternative is to leverage the parameter optimization procedure to precisely account for such noises in an implicit manner, given that the optimization occurs faster than the \ac{QC} drift in the device calibration. A second approach is to account for the errors by post-processing the measurements of the circuit through classical methods. 

\section{Quantum Approximate Optimization Algorithm (QAOA)}

\ac{QAOA} is a dynamic optimization method that leverages the potential of \ac{NISQ} devices to solve complex optimization problems first introduced by \citet{farhi2014quantum}. It operates by alternating between two types of quantum operations: problem-specific unitaries and mixer unitaries, applied iteratively to an initial quantum state. The algorithm aims to find the ground state of a cost Hamiltonian $H_C$, which encodes the problem to be solved. \ac{QAOA} leverages the principles of variational methods, where a cost function dependent on the quantum state is minimized to identify optimal solutions.

This section first describes combinatorial optimization in general and then introduces \ac{QAOA}. 

\subsection{Combinatorial Optimization}
Following \citet{farhi2014quantum}, we define a \ac{COP} as binary assignment decision problem with $n$ bits and $m$ clauses, where clauses represent constraints on subsets of bits. 
With the bitstring represented as $\mathbf{s}=s_1\dots s_n$, the objective function is specified as~\cite{farhi2014quantum}: 
\begin{equation}
    C(\mathbf{s})=\sum_{\alpha=1}^mC_\alpha(\mathbf{s}) \text{,}
\end{equation}
which corresponds to a binary assignment since $C_\alpha(\mathbf{s})=1$ if $\mathbf{s}$ satisfies $\alpha$, and 0 otherwise. The goal is to find a bitstring configuration so that the value of $C$ is maximized. While MAX-SAT (maximum satisfiability problem) searches for a string that achieves the Supremum of the objective function $C^*$, approximate optimization is tasked with finding a $\mathbf{s}^*$ for which the $C(\mathbf{s}^*)$ is close to $C^*$. \ac{QAOA} falls under the domain of approximate optimization.

\subsection{Mapping Combinatorial Optimization for Quantum Computation}

\label{subsec:mapping}
One can encode the binary assignment problem with variables $s_i\in\{-1, 1\}$, which allows the direct mapping to eigenvalues of the Pauli Z operator $\sigma^z = \left(\begin{smallmatrix}
1 & 0 \\
0 & -1
\end{smallmatrix}\right)$~\cite{khairy2020learning}.
The two eigenvalues of $\sigma^z$ are $\pm1$, which allows a direct mapping between $s_i$ (with values $\pm1$) and $\sigma^z$.
I.e., if $s_i=1$, it corresponds to the eigenstate $\ket{0}$ of $\sigma^z$ (with eigenvalue $+1$). Analogous for $s_1=-1$ with $\ket{1}$. 
Additionally, one usually encodes the clauses through a set of operations applied to a subset of qubits, and the cost through an aggregation function of such operators. 

\subsection{QAOA}
\label{subsec:qaoa}
Mathematically, on a high-level the \ac{QAOA} involves the following steps.

\fakeparagraph{1. Initialization}: 
The initial quantum state $\ket{\phi}$ is usually chosen to be a superposition of all possible states, such as the equal superposition state:
$
    \ket{\phi} = \frac{1}{\sqrt{2^n}} \sum_{z \in \{0, 1\}^n} \ket{z}
$
, where $z \in \{0, 1\}^n$ denotes any binary string of length $n$.
Another alternative representation is the uniform superposition of $n$ quantum states $\ket{\phi}=\ket{+}^{\otimes n}=\mathtt{H}^{\otimes n}\ket{0}^{\otimes n}$, which corresponds to applying Hadamard gates $\mathtt{H}$ on all qubits.  


\fakeparagraph{2. Alternating Unitaries}: The algorithm alternates between applying the problem unitary \( U(C, \gamma) = e^{-i \gamma H_C} \) and the mixer unitary \( U(M, \beta) = e^{-i \beta H_M}\) for a specified number of iterations \( p \). Here, \( H_C \) is the cost Hamiltonian corresponding to the objective function, and \( H_M \) is the mixer Hamiltonian, often chosen as a simple transverse field $H_M = \sum_{i=1}^n \sigma^x_i$, where \( \sigma^x_i \) is the Pauli-X operator acting on the \( i \)-th qubit. 

\ac{QAOA} applies a series of $k \in [p]$ alternating operators $e^{-i\beta_k H_M}$ and $e^{-i\gamma_k H_C}$ to prepare the variational quantum state $\ket{\phi(\boldsymbol{\beta}, \boldsymbol{\gamma})}$ as: 
\begin{equation}
    \ket{\phi(\boldsymbol{\beta}, \boldsymbol{\gamma})} = \prod_{j=1}^p U(M, \beta_j) U(C, \gamma_j) \ket{\phi} = e^{-i\beta_p H_M}e^{-i\gamma_p H_C} \dots e^{-i\beta_1 H_M}e^{-i\gamma_1 H_C} \ket{\phi} \text{,} 
\end{equation}
with a total of $2p$ variational parameters $\boldsymbol{\beta}, \boldsymbol{\gamma}$ that are optimized by a classical computer. 

\fakeparagraph{3. Cost Function Evaluation}
The cost function $C(\boldsymbol{\beta}, \boldsymbol{\gamma})$ to minimize is the expectation of the cost Hamiltonian $H_C$:
\begin{equation}
    C(\boldsymbol{\beta}, \boldsymbol{\gamma}) = \bra{\phi(\boldsymbol{\beta}, \boldsymbol{\gamma})}H_C\ket{\phi(\boldsymbol{\beta}, \boldsymbol{\gamma})} \text{.}
    \label{obj-func}
\end{equation}

\fakeparagraph{4. Optimization}: The parameters \( \boldsymbol{\gamma} \) and \( \boldsymbol{\beta} \) are optimized using classical optimization techniques to minimize \( C(\boldsymbol{\gamma}, \boldsymbol{\beta}) \). This iterative process aims to find the parameters that produce the lowest possible value of the cost function, thus approximating the ground state of \( H_C \).

\fakeparagraph{Relevance}
By alternating between problem-specific and mixer unitaries, the QAOA explores the solution space in a manner analogous to classical annealing processes. Additionally, it also harnesses the potential advantages of quantum superposition and entanglement for finding solutions to the problem at hand.
\ac{QAOA} has been identified in the literature as one of the most promising approaches for leveraging \ac{NISQ} devices for reaching quantum supremacy over classical computers~\cite{cerezo2021variational}.
In fact, \citet{lloyd2018quantum} shows how \ac{QAOA} is able to perform universal quantum computation. 
Moreover, \citet{morales2020universality} further prove the generality of \ac{QAOA} by showing is capacity to approximate any unitary.

\section{QAOA for Max-Cut Problem}

The MaxCut problem is a well-known combinatorial optimization problem in graph theory proven to be NP-Hard~\cite{commander2009maximum}. It involves partitioning the vertices of a given graph into two disjoint subsets such that the \emph{cut} of the graph is maximized -- i.e., maximize the sum of the weights of the edges between the two subsets. MaxCut is relevant in various fields, including computer science, operations research, and physics. In practical applications, it is used in network design, and circuit layout design. Its significance lies in its ability to model and solve real-world problems where division of resources or optimization of connectivity is crucial. 
QAOA is particularly effective for problems such as the Max-Cut and other constraint satisfaction problems, where the cost Hamiltonian \( H_C \) can be constructed to represent the problem's constraints and objective.

\subsection{Mapping MaxCut for \ac{QAOA}}
We represent the MaxCut \ac{COP} using the cost function $C$ for a given graph $G$ with $V$ vertices and $E$ edges $G=\langle V, E \rangle$ as:
$
C=\sum_{i,j \in V} w_{ij}\sigma^z_i\sigma^z_j \text{,}    
$
where $w_{ij}=1$ if $(i,j)\in E$ and $0$ otherwise~\cite{khairy2020learning}. The objective is to maximize this value. However, we apply negation transformation to convert it to a minimization problem. 

For \ac{QAOA} purposes, an edge between two nodes $i$ and $j$ in the graph includes a term in the Hamiltonian proportional to $\sigma^z_i\sigma^z_j$~\cite{khairy2020learning}, which becomes $1$ if the two qubits (corresponding to two vertices in the graph) are in the same state and $-1$ otherwise. This corresponds to the mapping described in Sec.~\ref{subsec:mapping} using the Pauli-Z operator.
Hence, the cost Hamiltonian for MaxCut becomes: $H_C = \sum_{(u, v) \in E} (\text{ZZ}(q_u, q_v))^\gamma$, where ZZ correspond to the ZZ interaction operation. For the mixer Hamiltonian $H_M$, we choose the simple traverse field operator as described in Sec.~\ref{subsec:qaoa}.

\subsection{Experimentation}
In this subsection we describe the dataset used, the implementation of the quantum circuit and the benchmarks used. We release our code, dataset and optimized circuit parameters under \url{https://github.com/DanielFPerez/VQA-for-MaxCut}. 

\fakeparagraph{Dataset} We consider graphs of 10 and 20 nodes. We generate 100 different graphs for each node configuration with random edges between the nodes according to the Erdős–Rényi model~\cite{erdos1959random} using the NetworkX python library~\cite{aric2008networkx}.

\begin{figure}
    \centering
    \subfloat[Graph to find a MaxCut for.\label{subfig:graph}]{%
       \includegraphics[width=0.25\textwidth]{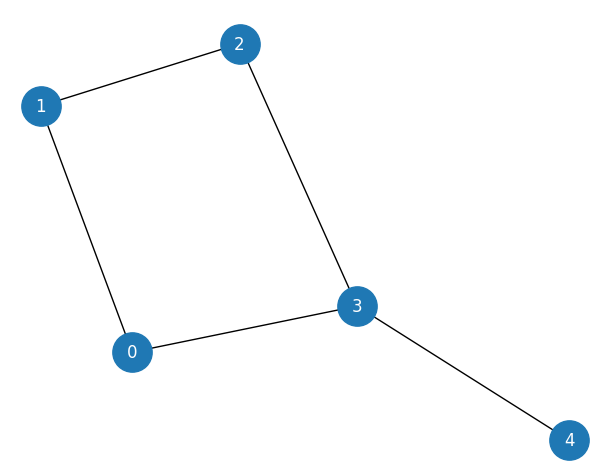}}
    \hfill
    \subfloat[Corresponding quantum circuit for the graph under (a). 
    \label{subfig:graph-circuit}]{%
        \includegraphics[width=0.7\textwidth]{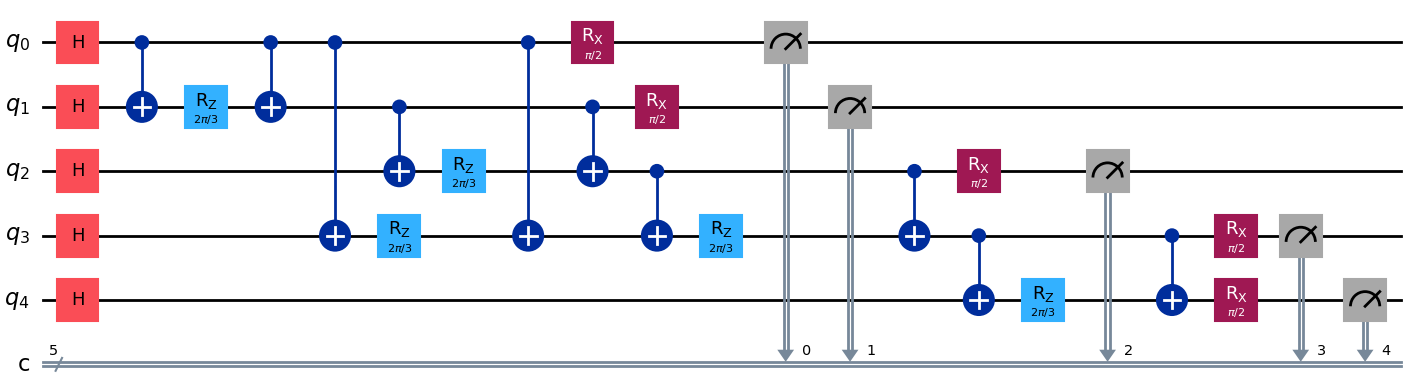}}
       
    \caption{When applying \ac{QAOA} to find the MaxCut for a given graph, a tailored quantum circuit must be build to encode the cost function for that specific graph. Hence, any changes to the graph require a new circuit to be optimized.} \label{fig:maxcut}
\end{figure}

\fakeparagraph{Circuit Implementation} For circuit design, we simulate the quantum circuit using Google's Cirq library~\cite{cirq_developers_2024_11398048}. We consider \ac{QAOA} circuits of different depths $p={1, 2, 3}$. For any given graph, one must construct a corresponding circuit that encodes the cost Hamiltonian $H_C$ due to its dependency on the graph edges. Hence, any changes to the connectivity of the graph implies generating and optimizing a new quantum circuit.
Fig.~\ref{fig:maxcut} visualizes a random graph with its corresponding \ac{QAOA} circuit for MaxCut. Note that in Fig.~\ref{subfig:graph-circuit}, the circuit implements the ZZ interaction as a sequence of CNOT-RZ$(\lambda)$-CNOT operations.

\fakeparagraph{Optimization} We choose two different optimizers: a gradient-based approach, and a non-gradient based approach. We implement the two types of optimizers with the SciPy library~\cite{virtanen2020scipy}. As \textbf{gradient-based} approach we choose the Broyden–Fletcher–Goldfarb–Shanno (BFGS) algorithm~\cite{nocedal1999numerical}, a well-used iterative method for nonlinear optimization, which updates an approximation to the Hessian matrix to improve the convergence rate towards the minimum of a function. It combines gradient descent with quasi-Newton methods to efficiently handle large-scale optimization tasks. For evaluating the gradient, we use the parameter-shift rule~\cite{schuld2019parametershift, mitarai2018quantum}. 
As a \textbf{non-gradient} method, we choose the Nelder-Mead method~\cite{nelder1965simplex} due to its use in the Quantum Variational Eigensolver~\cite{peruzzo2014variational} and related \ac{VQA} tasks. It is a derivative-free heuristic algorithm for finding the minimum of a function by iteratively refining a simplex\footnote{For an objective function with $n$ variables, a simplex is a convex hull of $n+1$ vertices, each vertex is of dimension $\mathbb{R}^n$.} of points in the search space based on geometric transformations (reflection, expansion, contraction, and shrinking operations). At each iteration, the objective function is evaluated on each vertex of the simplex, and then a type of transformation is chosen. 

\fakeparagraph{Benchmark} We compare the results of the \ac{QAOA} with a classical method: the one-exchange algorithm implemented in NetworkX for MaxCut~\cite{aric2008networkx}. It is a greedy local search approximation algorithm using the one-exchange heuristic: it starts with an initial partition of the vertices and iteratively attempts to improve the partition by considering the effect of moving a single vertex from one subset to the other.

\subsection{Results}

\begin{figure}
    \centering
    \subfloat[\textbf{10-nodes Non-Gradient-based} MaxCut for each graph.\label{subfig:10-node-nongrad-indivgraph}]{%
       \includegraphics[clip, trim=3.5cm 0cm 3.5cm 0cm, width=0.495\textwidth]{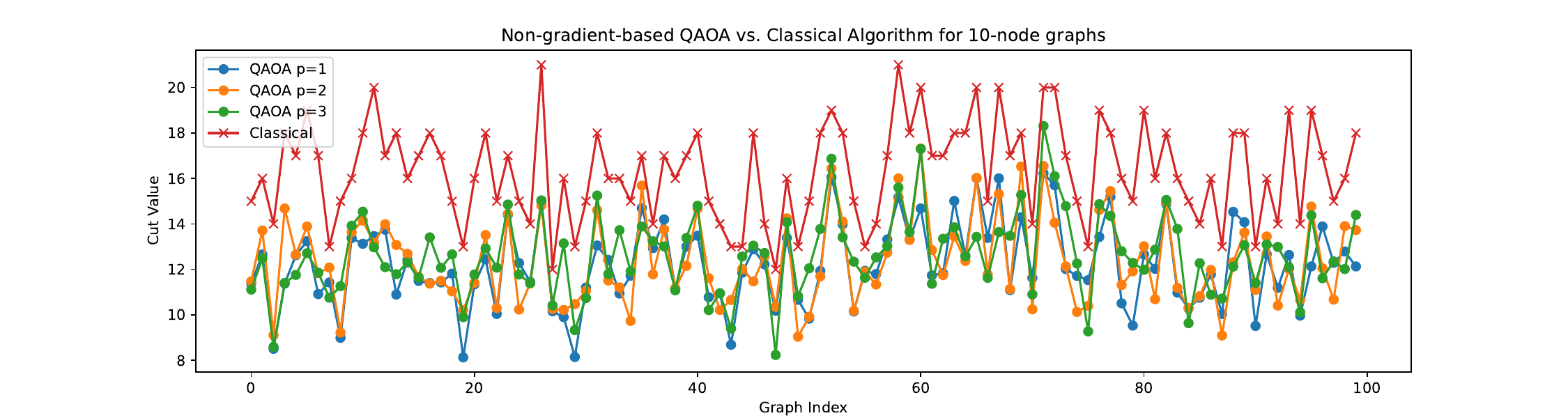}}
    \hfill
    \subfloat[\textbf{20-nodes Non-Gradient-based} MaxCut for each graph.\label{subfig:20-node-nongrad-indivgraph}]{%
       \includegraphics[clip, trim=3.5cm 0cm 3.5cm 0cm, width=0.495\textwidth]{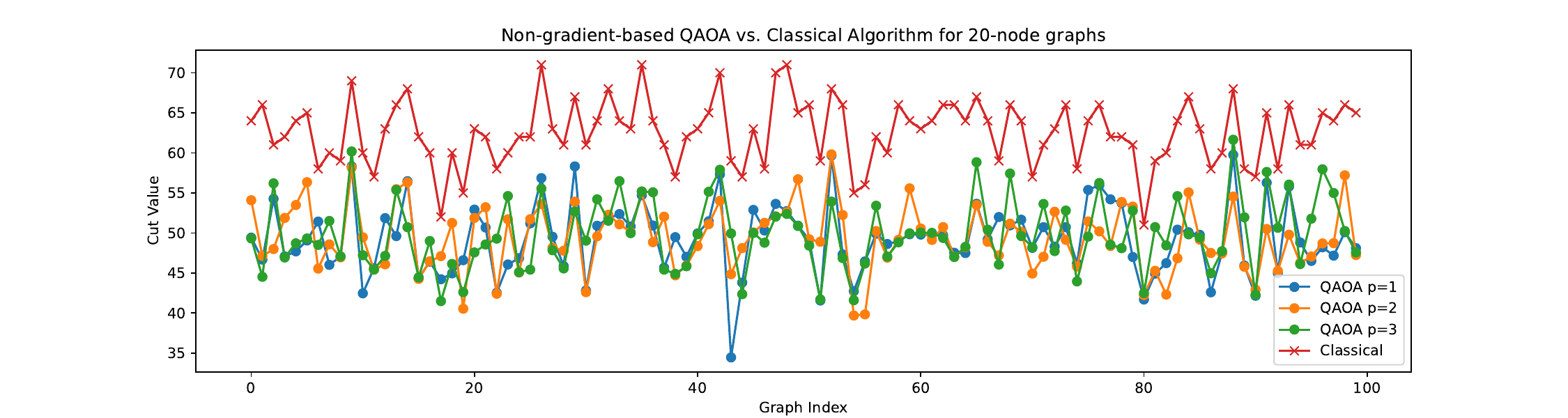}}
    \\
    \subfloat[\textbf{10-nodes Gradient-based} MaxCut for each graph.\label{subfig:10-node-grad-indivgraph}]{%
       \includegraphics[clip, trim=3.5cm 0cm 3.5cm 0cm, width=0.495\textwidth]{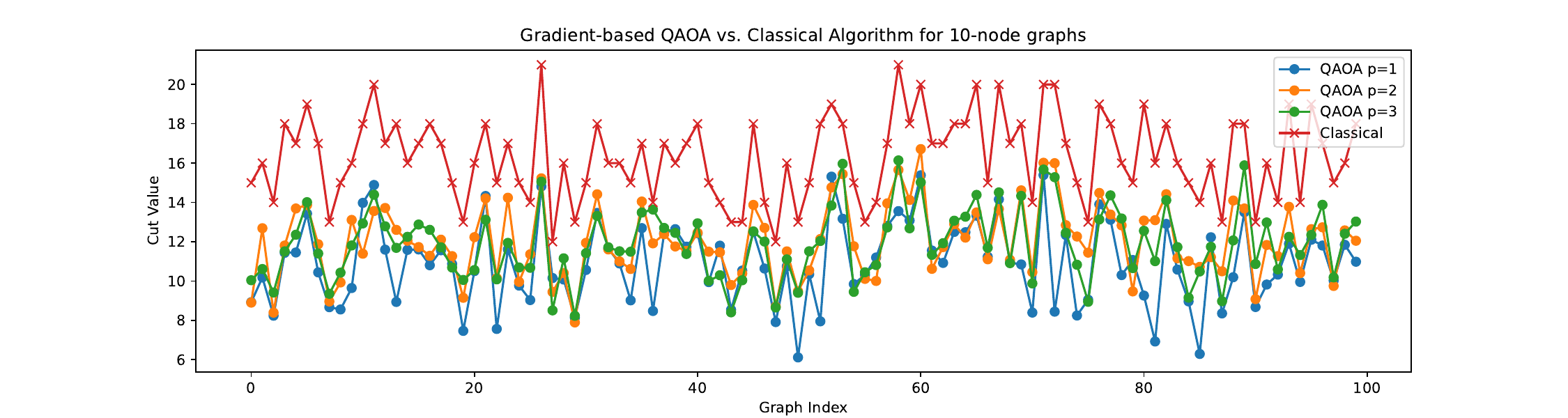}}
    \hfill
    \subfloat[\textbf{20-nodes Gradient-based} MaxCut for each graph.\label{subfig:20-node-grad-indivgraph}]{%
       \includegraphics[clip, trim=3.5cm 0cm 3.5cm 0cm, width=0.495\textwidth]{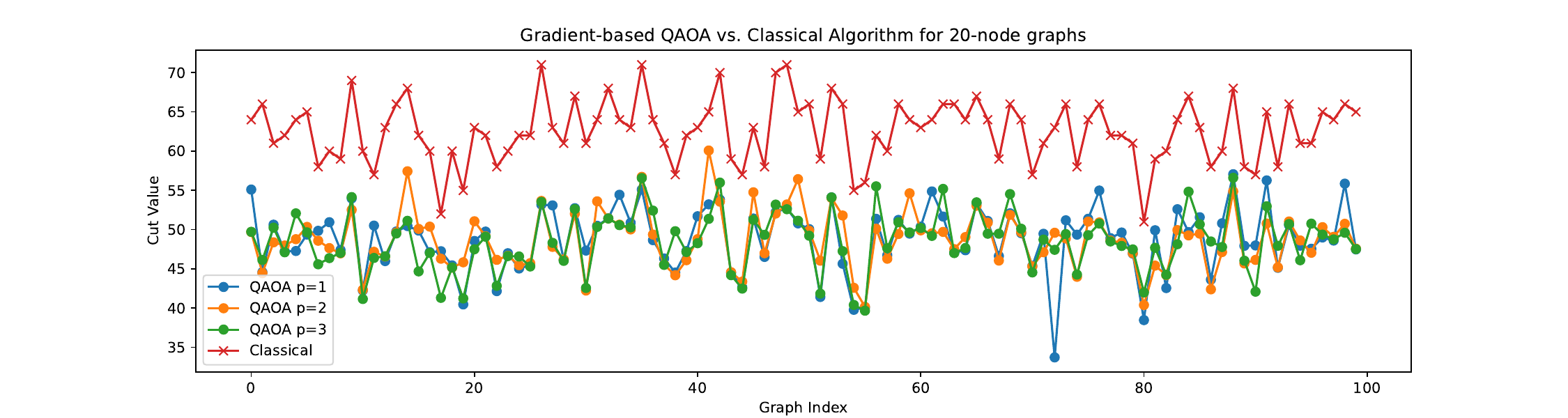}}
    \\
    
    \subfloat[\textbf{10-nodes Non-Gradient-based} average MaxCut normalized w.r.t. classical method. 
    \label{subfig:10-node-nongrad-bars}]{%
        \includegraphics[width=0.24\textwidth]{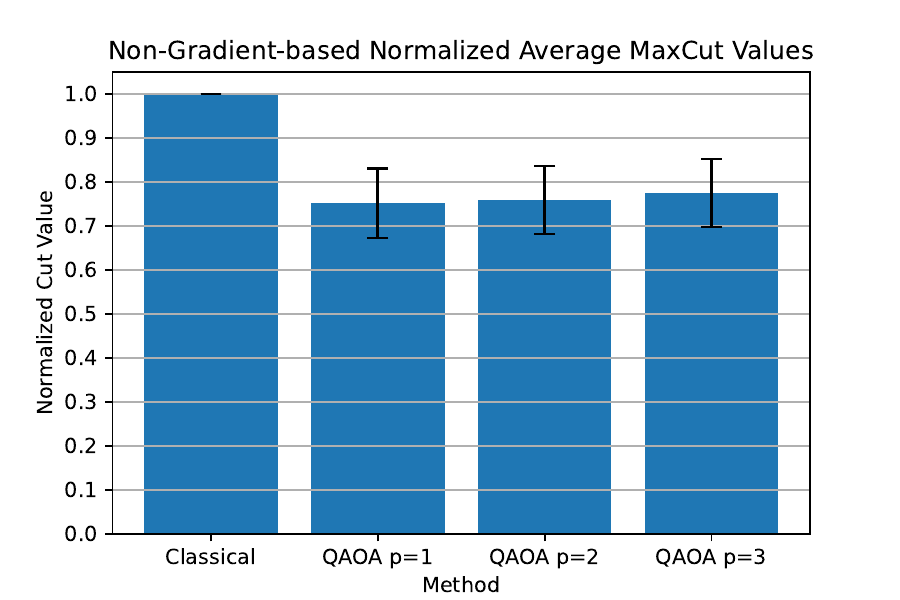}}
    \hfill
    \subfloat[\textbf{10-nodes Gradient-based} average MaxCut normalized w.r.t. classical method. 
    \label{subfig:10-node-grad-bars}]{%
        \includegraphics[width=0.24\textwidth]{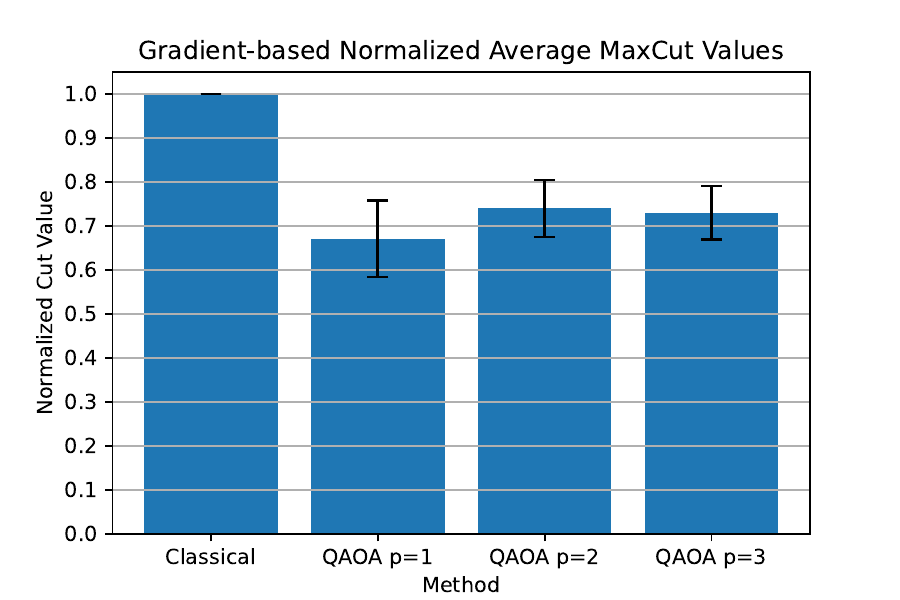}}
    \hfill
    \subfloat[\textbf{20-nodes Non-Gradient-based} average MaxCut normalized w.r.t. classical method. 
    \label{subfig:20-node-nongrad-bars}]{%
        \includegraphics[width=0.24\textwidth]{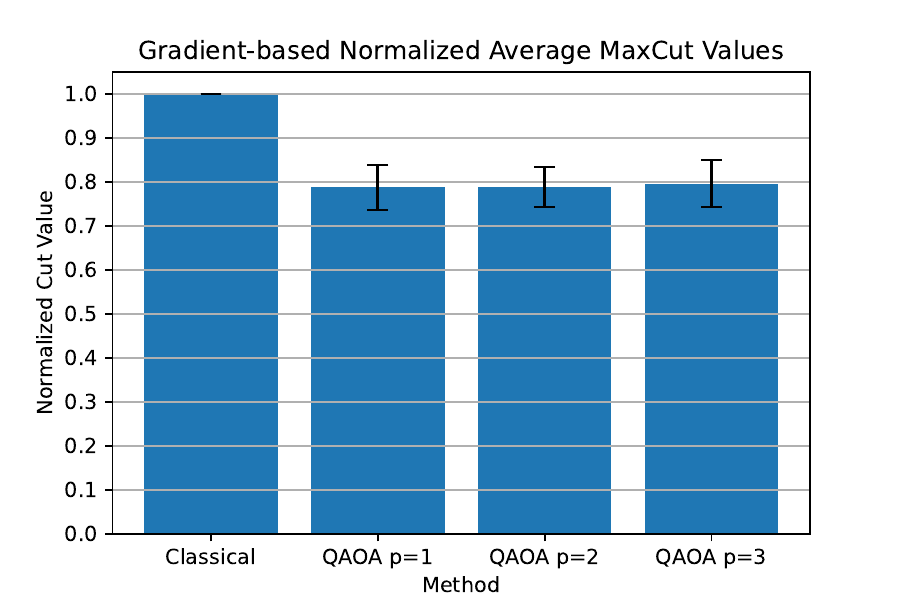}}
    \hfill
     \subfloat[\textbf{20-nodes Gradient-based} average MaxCut normalized w.r.t. classical method. 
    \label{subfig:20-node-grad-bars}]{%
        \includegraphics[width=0.24\textwidth]{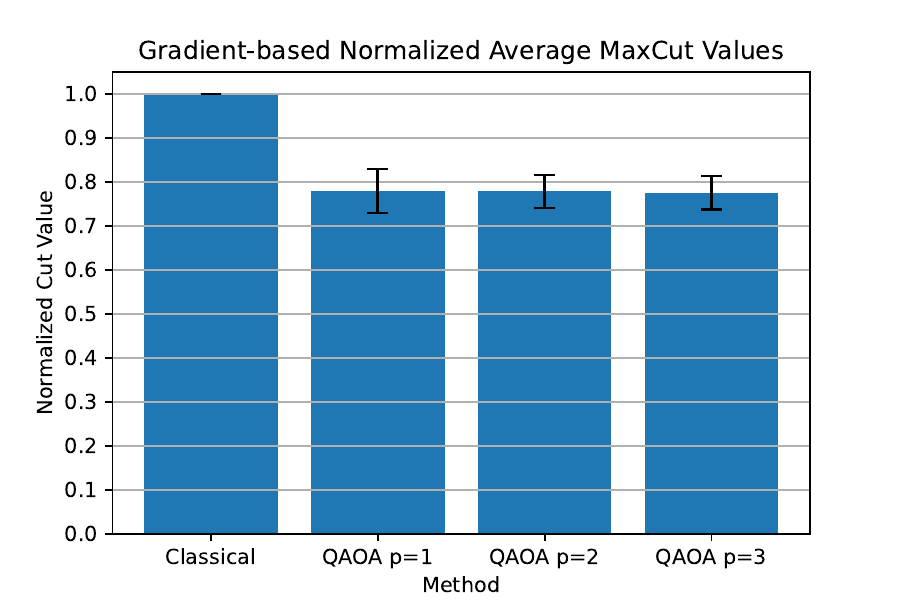}}
        
    \caption{Overview of results comparing (gradient-, and non-gradient based) \ac{QAOA} against a classical method for 100 Erdős–Rényi graphs of 10 and 20 nodes. Vertical lines represent the standard deviation.} \label{fig:results}
\end{figure}

The experimentation results are depicted in Fig.~\ref{fig:results} for both 10-node and 20-node graphs. In general, the \ac{QAOA} delivers descent MaxCut values, especially for a circuit depth of $p=2$ using the gradient-based approach for both graph sizes considered. More specifically, \ac{QAOA} approaches seem to achieve between $0.7$ to $0.8$ of the MaxCut values of their classical counterpart, even without cumbersome parameter initialization or more sophisticated optimization techniques. However, the classical method outperforms the \ac{QAOA} approaches in the experiments considered.

As one can observe in  Figs.~\ref{subfig:10-node-nongrad-indivgraph}-\ref{subfig:20-node-grad-indivgraph} when comparing the MaxCut of each individual graph for different p-values, the \ac{QAOA} follows a similar qualitative profile in comparison to its classical counterpart. Notably, the classical counterpart always outperforms the \ac{QAOA} method, despite increasing circuit depth $p$ and type of optimizer used. For the 10 nodes graphs, Figs~\ref{subfig:10-node-nongrad-indivgraph} and \ref{subfig:10-node-grad-indivgraph} show how $p=1$ delivers the lowest MaxCut values observed for some of the graphs, which is more prominent for the gradient-based approach considered. 
This phenomenon is also present for the 20-node graphs in Figs.~\ref{subfig:20-node-nongrad-indivgraph} and~\ref{subfig:20-node-grad-indivgraph}, but with no significant difference based on the optimizers.
Moreover, one observes a similar profile for both $p=2$ and $p=3$ in Figs.~\ref{subfig:10-node-nongrad-indivgraph}-\ref{subfig:20-node-grad-indivgraph}, which suggests that there is no significant gain in increasing the circuit depth beyond $p=2$. An interesting finding when observing the individual MaxCut values for all graphs considered, is that on some occasions, the a circuit depth of $p=1$ outperforms deeper circuits of $p=\{2, 3\}$.

Figs.~\ref{subfig:10-node-nongrad-bars}-~\ref{subfig:20-node-grad-bars} depict the average MaxCut values across the 100 graphs normalized by the average MaxCut value of the classical method for different $p$ values and optimizer used. One observes that as the graph size increases, the achievable average MaxCut increases regardless of the optimizer used. The maximum average MaxCut value observed for \ac{QAOA} is $0.8$ for the 20 node graphs and non-gradient optimizer with depth $p=3$ in Fig.~\ref{subfig:20-node-nongrad-bars}. However, the $p=2$ \ac{QAOA} achieves better results than $p=1$ for the 10 node graphs (Figs.~\ref{subfig:10-node-nongrad-bars} and~\ref{subfig:10-node-grad-bars}), and similar results than $p=3$ overall, but with the advantage of a slight reduction of the standard deviation and a more simple circuit structure complexity. 

\begin{table}[h!]
    \centering
    \begin{tabular}{ccccccc}
        \toprule
        & \multicolumn{3}{c}{Gradient-based} & \multicolumn{3}{c}{Non-gradient based} \\
        \cmidrule(r){2-4} \cmidrule(l){5-7}
        \ac{QAOA} Circuit Depth $p$ & $p=1$ & $p=2$ & $p=3$ & $p=1$ & $p=2$ & $p=3$ \\
        \midrule
        Average runtime [sec] & \textbf{2.1} & \textbf{4.12} & \textbf{7.9} & 8.2 & 20.39 & 36.7 \\
        Std-dev. [sec] & 1.18 & 1.31 & 2.93 & 0.56 & 1.32 & 2.47 \\
        \bottomrule
    \end{tabular}
    \caption{\textbf{Gradient-based  optimization performs faster}. Average runtime comparison of \ac{QAOA} circuit optimization for \textbf{10-node} graphs of both Gradient-based and Non-gradient based methods considering different circuit depths $p$.}
    \label{table:runtime}
\end{table}

While there was no significant difference between the optimizers used in terms of MaxCut values achieved, there is a significant difference in terms of runtimes. We cache the average runtime values for the 10-node topologies for different circuit depths $p$, together with their standard deviation. The results are shown in Table~\ref{table:runtime}. It depicts how the gradient-based optimizer performs much faster than the non-gradient counterpart.

\section{Conclusion}

In this paper, we start by identifying the most promising quantum computing algorithms in the \ac{NISQ} era, Variational Quantum Algorithms. We then describe a specific Ansatz and algorithm specifically designed to tackle combinatorial optimization problems (COP): the Quantum Alternating Optimization Algorithm (\ac{QAOA}). \ac{QAOA} has been identified in the literature as a potential approach to perform universal quantum computations, with the capacity to approximate any unitary. Hence, it is of high relevance to further study its potential application to solve \acp{COP}. Finally, we apply \ac{QAOA} for solving the MaxCut problem of graphs up to 10-nodes and 20-nodes. We show how \ac{QAOA} is able to achieve in average $80\%$ of the MaxCut value of the classical benchmark considered, even with shallow \ac{QAOA} circuits and without an extensive hyper-parameter optimization.  
While shallow circuit designs and simple optimization configuration allow us to effectively apply \ac{QAOA} for MaxCut, further analysis is required to truly show the potential superiority of \ac{QAOA} over classical methods.

\begin{acks}
This work was financially supported by the Swedish Foundation for Strategic Research. 
\end{acks}

\bibliographystyle{ACM-Reference-Format}
\bibliography{refs}

\input{acronyms}

\end{document}

%% file: acronyms.tex
\acrodef{TSCH}{Time-Slotted Channel Hopping}
\acrodef{RF}{Radio Frequency}
\acrodefindefinite{RF}{an}{a}
\acrodef{LNA}{Low-Noise Amplifier}
\acrodefindefinite{LNA}{an}{a}
\acrodef{LO}{Local Oscillator}
\acrodefindefinite{LO}{an}{a}
\acrodef{ADC}{Analog-to-Digital Converter}
\acrodef{IF}{Intermediate Frequency}
\acrodef{CMOS}{Complementary Metal-Oxide-Semiconductor}
\acrodef{DBP}{Digital Baseband Processor}
\acrodef{DSSS}{Direct Sequence Spread Spectrum}
\acrodef{ASK}{Amplitude Shift Keying}
\acrodef{MSK}{Minimum Shift Keying}
\acrodefindefinite{MSK}{an}{a}
\acrodef{FSK}{Frequency Shift Keying}
\acrodefindefinite{FSK}{an}{a}
\acrodef{oqpsk}[O-QPSK]{Offset-Quadrature Phase Shift Keying}
\acrodef{BPF}{Band-Pass Filter}
\acrodef{PRR}{Packet Reception Ratio}
\acrodef{SDR}{Software Defined Radio}
\acrodefindefinite{SDR}{an}{a}
\acrodef{IC}{Integrated Circuit}
\acrodef{WSN}{Wireless Sensor Networks}
\acrodef{CCA}{Clear Channel Assessment}
\acrodef{MAC}{Medium Access Control}
\acrodef{IoT}{Internet of Things}

\acrodef{NIC}{Network Interface Card}
\acrodef{SmartNIC}{Smart Network Interface Card}
\acrodef{HOL}{Head-of-Line blocking}
\acrodef{RMT}{Reconfigurable Match Tables}
\acrodef{SLO}{Service-Level Objective}
\acrodef{QoS}{Quality of Service}
\acrodef{QoE}{Quality of Experience}
\acrodef{IDS/IPS}{Intrusion Detection and Prevention Systems}
\acrodef{FPGA}{Field-Programmable Gate Array}
\acrodef{RTT}{Round-Trip Time}
\acrodef{RMA}{Remote Memory Access}
\acrodef{RDMA}{Remote Direct Memory Access}

\acrodef{DAG}{Direct Acyclic Graph}
\acrodef{COP}{Combinatorial Optimization Problem}
\acrodef{ML}{Machine Learning}
\acrodef{NN}{Neural Network}
\acrodefindefinite{ML}{an}{a}
\acrodef{GRL}{Graph Representation Learning}
\acrodef{GNN}{Graph Neural Networks}
\acrodef{BN}{Bayesian Network}
\acrodef{RNN}{Recurrent Neural Network}
\acrodef{NMT}{Neural Machine Translation}
\acrodef{RL}{Reinforcement Learning}
\acrodef{DRL}{Deep Reinforcement Learning}
\acrodef{SSL}{Self-Supervised Learning}
\acrodef{seq2seq}{Sequence-to-Sequence}
\acrodef{LSTM}{Long Short-Term Memory}
\acrodef{CNN}{Convolutional Neural Network}
\acrodef{GBT}{Gradient Boosting Trees}
\acrodef{BNN}{Binary Neural Network}
\acrodef{KDE}{Kernel Density Estimator}

\acrodef{QC}{Quantum Computer}
\acrodef{PQC}{Parameterized Quantum Circuit}
\acrodef{VQA}{Variational Quantum Algorithm}
\acrodef{VQC}{Variational Quantum Circuit}
\acrodef{NISQ}{Noisy Intermediate-Scale Quantum}
\acrodef{QAOA}{Quantum Approximate Optimization Algorithm}
\acrodef{BP}{Barren Plateaus}
\acrodef{QML}{Quantum Machine Learning}
\acrodef{QNN}{Quantum Neural Network}
\acrodef{QCNN}{Quantum Convolutional Neural Network}